\documentclass[11pt]{article}
\usepackage{moriond,epsfig}
\newcommand{\beq}{\begin{eqnarray}}
\newcommand{\eeq}{\end{eqnarray}}

\newcommand{\bmp}{\noindent\begin{minipage}{16cm}}
\newcommand{\emp}{\end{minipage}\vskip 7mm} 

\usepackage{graphicx}
\usepackage{dcolumn}
\usepackage{bm}
\usepackage{amsmath}
\usepackage{amsfonts}
\usepackage{bbm}
\usepackage{subfigure}

\def\be{\begin{equation}}
\def\ee{\end{equation}}
\def\bea{\begin{eqnarray}}
\def\eea{\end{eqnarray}}

\begin{document}
\vspace*{4cm}
\title{Dynamical Electroweak Symmetry Breaking}

\author{ F. Sannino }

\address{The Niels Bohr Institute, Blegdamsvej 17, Copenhagen \O, DK-2100, Denmark. }

\maketitle\abstracts{Dynamical breaking of the electroweak theory, i.e. technicolor, is
an intriguing extension of the Standard Model. Recently new models
have been proposed featuring walking dynamics for a very low number
of techniflavors. These technicolor extensions are not ruled out by current
precision measurements. Here I first motivate the idea of dynamical
electroweak symmetry breaking and then summarize some of the
properties of the recent models and their possible cosmological
implications.}

\section{Introduction}
The Standard Model (SM) of particle interactions is a low energy
effective theory valid up to a cutoff scale $\Lambda$. One of the
reasons behind the phenomenological success of the SM is that most
of the physical obeservables depend only logarithmically on
$\Lambda$. There is only one operator in the SM depending strongly
on this fundamental scale, i.e. the mass squared operator of the
Higgs. There are two problems associated  with such an operator: 1)
It is unnaturally sensitive to the scale $\Lambda$ which can be
taken to be the highest scale in the game, i.e. the Plank mass. 2)
Even if we set the Higgs mass operator to zero, at tree level, it will be
regenerated by quantum corrections.

Different resolutions of these problem have been proposed. Here I
will just mention the time-honored ones: i)  supersymmetric
generalizations of the SM ii) dynamical breaking of the electroweak
theory.

Combining the problems above with the fact that the SM alone neither
accounts for the experimentally observed dark matter\footnote{Here I
am barring exotic solutions to the dark matter problem stemming from
the SM augmented with Gravity.} in the Universe nor explains why the
neutron electric dipole moment is so unnaturally small we arrive at
the conclusion that it is {\it not so standard}  after all. Here I
will focus on the Higgs sector. Perhaps the first question to ask
is: Have we observed a Higgs-type mechanism in Nature?

Ordinary Superconductivity  (SC) is a noble example. In the first
figure, made by two slides, I summarize the key features that SC and
Electroweak Symmetry Breaking (ESB) have in common. SC and ESB  are
both an example of a screening effect. One can define a macroscopic
wave function $\psi$ in SC with $|\psi|^2 = n_c=n_s/2$,  $n_c$ the
number of Cooper pairs and $n_s$ the number of SC electrons.  This
wave function can be mapped into the Higgs wave function whose
square evaluated on the ground states $|\phi|=v^2/2$ sets the scale
of ESM, i.e. $v$ hundreds of GeV. {}For the few and even fewer
attentive readers who spotted the fact that the SC wave function
has different units than the Higgs wave function the reason is that
$\psi$ emerges in a nonrelativistic framework while $\phi$ is for a
relativistic one. However differences in units disappear when
comparing the Meissner static mass of the photon $M$ with the typical SM
gauge boson mass $M_W$. In the figure $q=-2e$ is the
electric charge of the Cooper pair and $m=2m_e$ is its mass. One can
then compare the typical screening lengths in the two cases.
\begin{figure}[!tbp]
  \begin{center}
    \mbox{
      \subfigure{\resizebox{!}{6cm}{\includegraphics{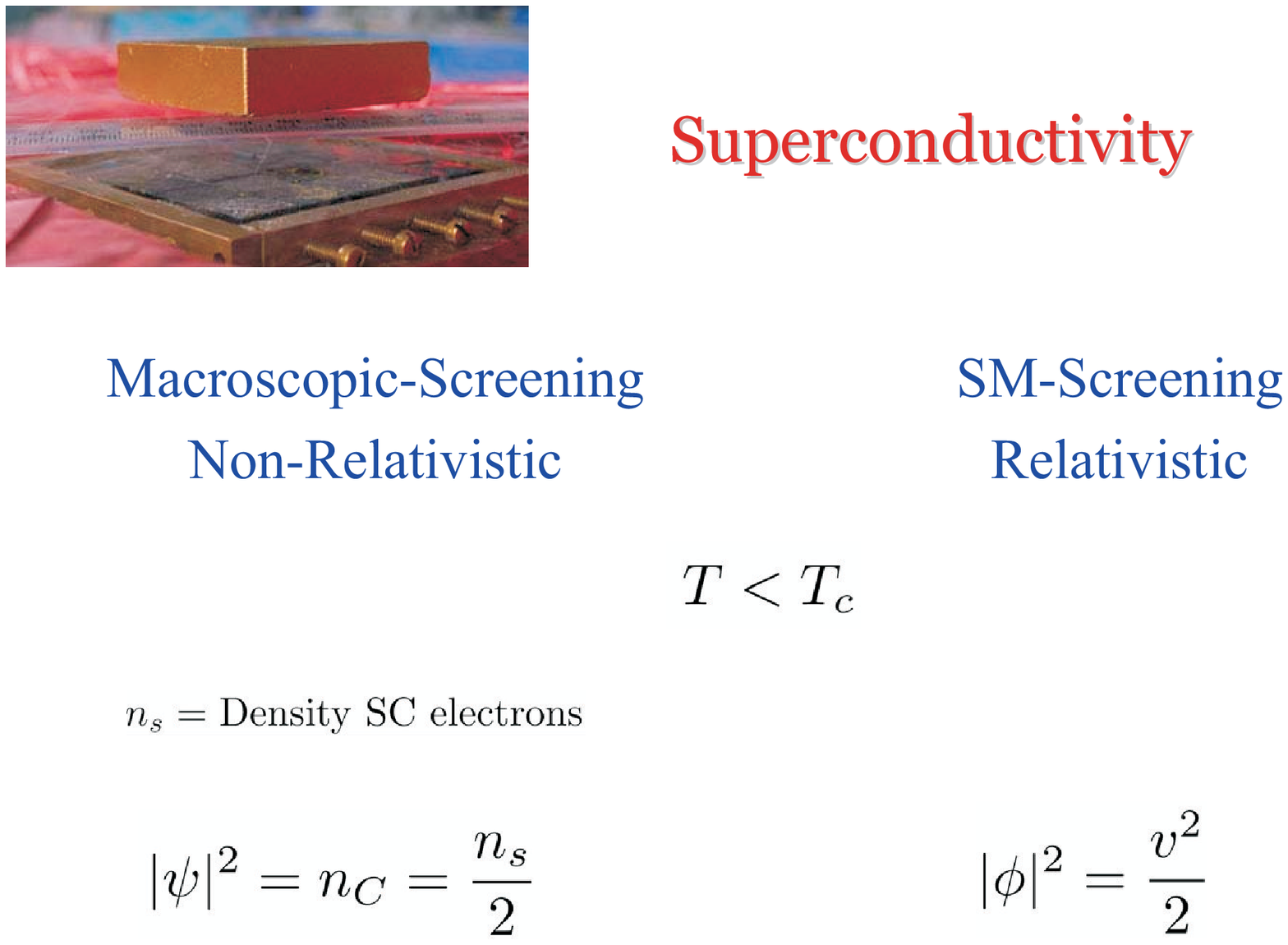}}} \hskip -1cm
      \subfigure{\resizebox{!}{6cm}{\includegraphics{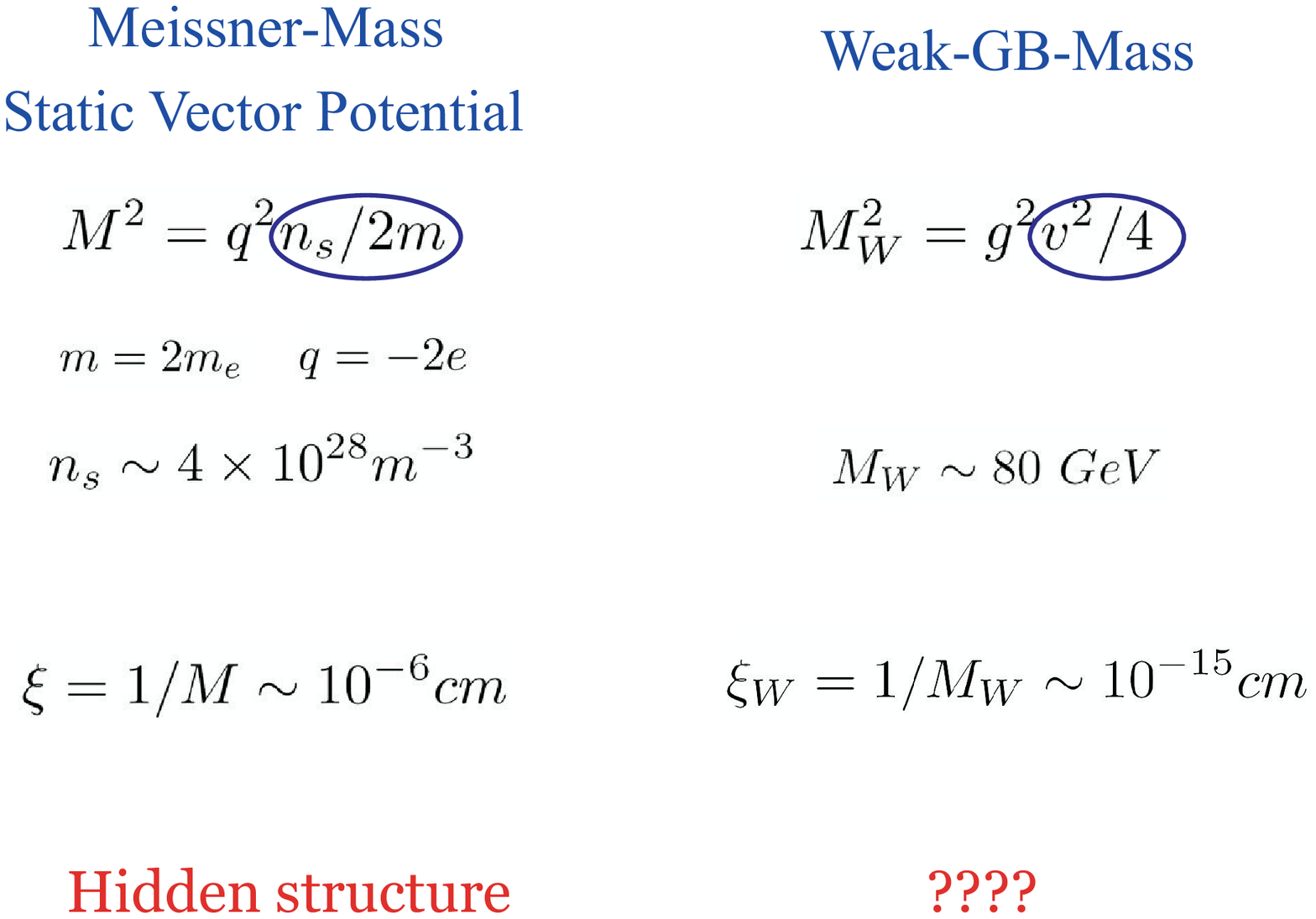}}}
      }
    \caption{Two slides summarizing common features of Superconductivity and Electroweak Symmetry Breaking. }
    \label{Fig1}
  \end{center}
\end{figure}
Having constructed the superconductive material in a lab we know
that the wave function $\psi$ is {\it not} a fundamental object but
rather a low energy description of something more fundamental. In
the SM we do not {\it yet} know the mechanism behind the ESB but it
might very well be dynamical as it is in SC. We will pursue
this idea here. By simply admitting that the Higgs wave function is not
associated to a fundamental field but is a low energy effective
description of a more fundamental theory valid up to a scale above
which we will be able to resolve its constituents one solves the
first two problems mentioned in the opening of this section. The highest scale in the
game is no longer Planck but rather the TeV scale.

We hence postulate the presence of a new strong force driving ESB.
Earlier attempts using QCD-like technicolor \cite{TC} have been
ruled out by precision measurements \cite{Peskin:1990zt}. Besides,
one has also to face the problem of mass generation which
is provided by extended technicolor (ETC) interactions and thus
leads to large flavor changing neutral currents. Recently, it has
been shown that one can construct viable theories explaining the
breaking of the electroweak theory dynamically
\cite{Sannino:2004qp,Dietrich:2005jn,Evans:2005pu}
while not being at odds with electroweak precision measurements. In
the recently proposed theories technimatter transforms according to
a higher dimensional representation of the new gauge group. By
direct comparison with data it turns out that the preferred
representation is the two-index symmetric \cite{Sannino:2004qp}. The
simplest theory of this kind is a two technicolor theory. In this
case the two-index symmetric representation coincides with the
adjoint\footnote{The use of higher dimensional representations for walking technicolor theories was suggested first in \cite{Lane:1989ej}.}. Remarkably, these theories are already near conformal for a
very small number of techniflavors. Further properties of higher
dimensional representations have also been explored in
\cite{Christensen:2005cb}.  In
\cite{Dietrich:2005jn,Evans:2005pu} the reader can
find a summary of a number of salient properties of the new
technicolor theories as well as a comprehensive review of the
walking properties with an exhaustive list of important references. We also note
that near the conformal window
\cite{Appelquist:1998xf,Sundrum:1991rf} one of the relevant
electroweak parameters ($S$) is smaller than expected in
perturbation theory. This observation is further supported by other very
recent analyses \cite{Hong:2006si,Harada:2005ru}.

\section{The Minimal Walking Model}
The new dynamical sector underlying the Higgs mechanism we consider
is an $SU(2)$ technicolor gauge group with two adjoint
technifermions. The theory is asymptotically free if the number of
flavors $N_f < 2.75$. The critical value of the number of flavors needed to reach the
infrared fixed point value is $N_f^c \simeq 2.075$
\cite{Sannino:2004qp,Evans:2005pu}.
We expect that the theory will enter a conformal regime before the
coupling rises above the critical value triggering the formation of
a fermion condensate. Hence, a $N_f=2$ theory is sufficiently close
to the critical number of flavors $N_f^c$. This makes it a perfect
candidate for a walking technicolor theory.

The two adjoint fermions may be written as \beq T_L^a=\left(
\begin{array}{c} U^{a} \\D^{a} \end{array}\right)_L , \qquad U_R^a \
, \quad D_R^a \ ,  \qquad a=1,2,3 \ ,\eeq with $a$ the adjoint color
index of $SU(2)$. The left fields are arranged in three doublets of
the $SU(2)_L$ weak interactions in the standard fashion. The
condensate is $\langle \bar{U}U + \bar{D}D \rangle$ which breaks
correctly the electroweak symmetry.

The model described so far suffers from the Witten topological
anomaly \cite{Witten:fp}. We can avoid this problem by adding a new weak doublet uncharged
under technicolor \cite{Dietrich:2005jn}. Our additional matter content resembles
a copy of a SM fermion family with quarks (here
transforming in the adjoint of $SU(2)$) and the following lepton
doublet \beq {\cal L}_L = \left( \begin{array}{c} N \\ E \end{array}
\right)_L , \qquad N_R \ ,~E_R \ . \eeq
Gauge anomaly cancellations do not fix uniquely the hypercharge for the additional matter.
In \cite{Evans:2005pu}, the SM-like
 hypercharge has been investigated in the context of an extended
technicolor theory. Another interesting choice for the hypercharge
has been investigated from the point of view of the electroweak
precision measurements, in \cite{Dietrich:2005jn}.
In that case
\begin{eqnarray}
Q(U)=1 \ , \quad Q(D)=0 \ , \quad Q(N)=-1 \ , \quad {\rm and }\quad
Q(E)=-2 \ .\end{eqnarray} Notice that with this particular hypercharge
assignment, the technidown $D$ is electrically neutral. Independently on the hypercharge assignment,
after spontaneous symmetry breaking we have nine Goldstone bosons. Three will become the longitudinal components of the massive vector
bosons and the other six will acquire mass via new interactions and
carry technibaryon number. The low energy effective theories have
been constructed in \cite{Gudnason:2006ug}.

\begin{figure}[htbp]
  \centerline{\epsfig{file=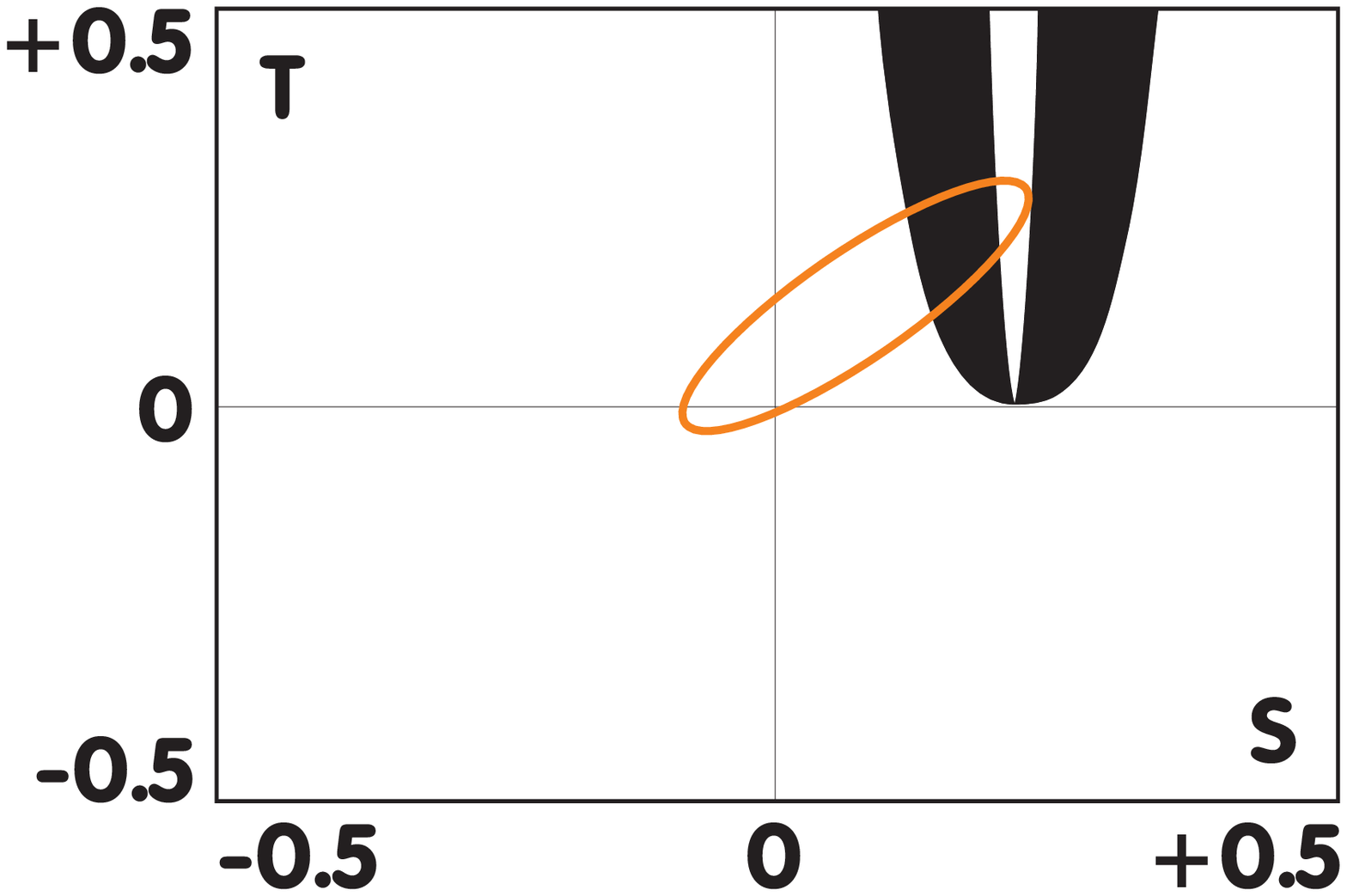,height=3.5cm}\hspace*{4mm}
              \epsfig{file=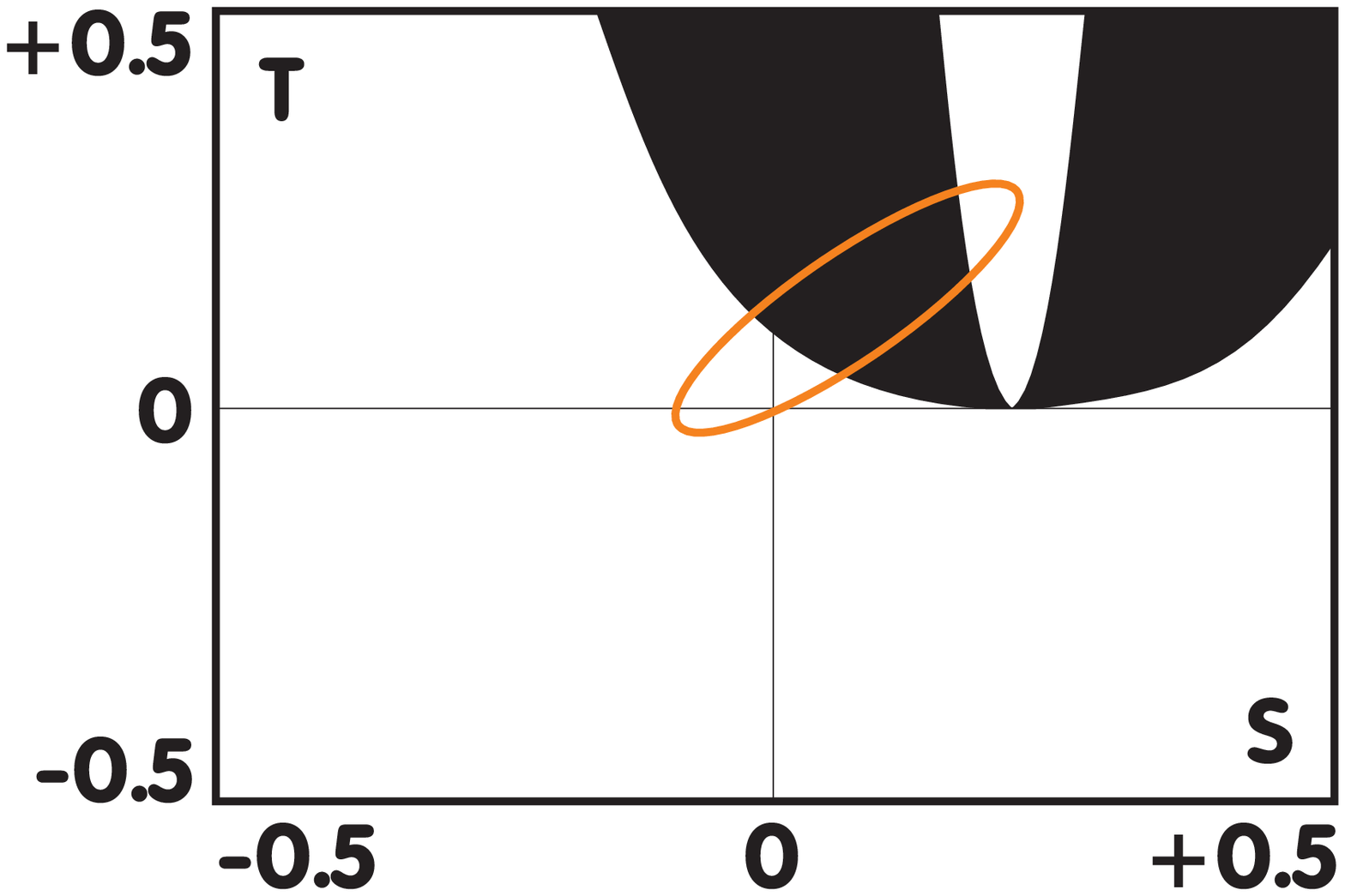,height=3.5cm}}
  \caption{Left Panel: The black shaded parabolic area corresponds to the accessible range of S and T for the extra neutrino and extra electron for masses from $m_Z$ to $10 m_Z$.
The perturbative estimate for the contribution to S from
techniquarks equals $1/2\pi$. The ellipse is the $90$\% confidence
level contour for the global fit to the electroweak precision data
with U kept at $0$. The contour is for the reference Higgs mass of
$m_H = 150$ GeV. Right Panel: Here the plot is obtained with a
larger value of the hypercharge choice, according to which one of
the two fermions is doubly charged and the other is singly charged
under the electromagnetic interactions.}
    \label{comparison}
\end{figure}

 A possible feature of these theories is that the resulting composite Higgs can be light with a
mass of the order of $150$~GeV. The phenomenology of these theories
leads to interesting signatures \cite{Zerwekh:2005wh}. It is
instructive to compare the present model with the new precision
measurements \cite{Dietrich:2005jn}. In figure \ref{comparison}
the ellipse corresponds to the one sigma contour in the T$-$S plane.
The black area bounded by parabolas corresponds to the region in the
T$-$S plane obtained when varying the Dirac masses of the
two new leptons. The point at T$=0$ where the inner parabola meets
the S axis corresponds to the contribution due solely to the
technicolor theory. The electroweak parameters are computed
perturbatively. {}Fortunately for walking technicolor theories the
nonperturbative corrections further reduce the S parameter
contribution \cite{Appelquist:1998xf} and hence
our estimates are expected to be rather conservative.
The figure clearly shows that the walking technicolor type theories
are viable models for dynamical breaking of the electroweak
symmetry. The central experimental values for $S$ and $T$
are respectively $S=+0.07 \pm 0.10$ and $T=+0.13\pm 0.10$.

\subsection{The Dark-Side \label{secdarkside}}

\begin{figure}[!tbp]
 \begin{center}
    \mbox{
      \subfigure{\resizebox{!}{4.8cm}{\includegraphics{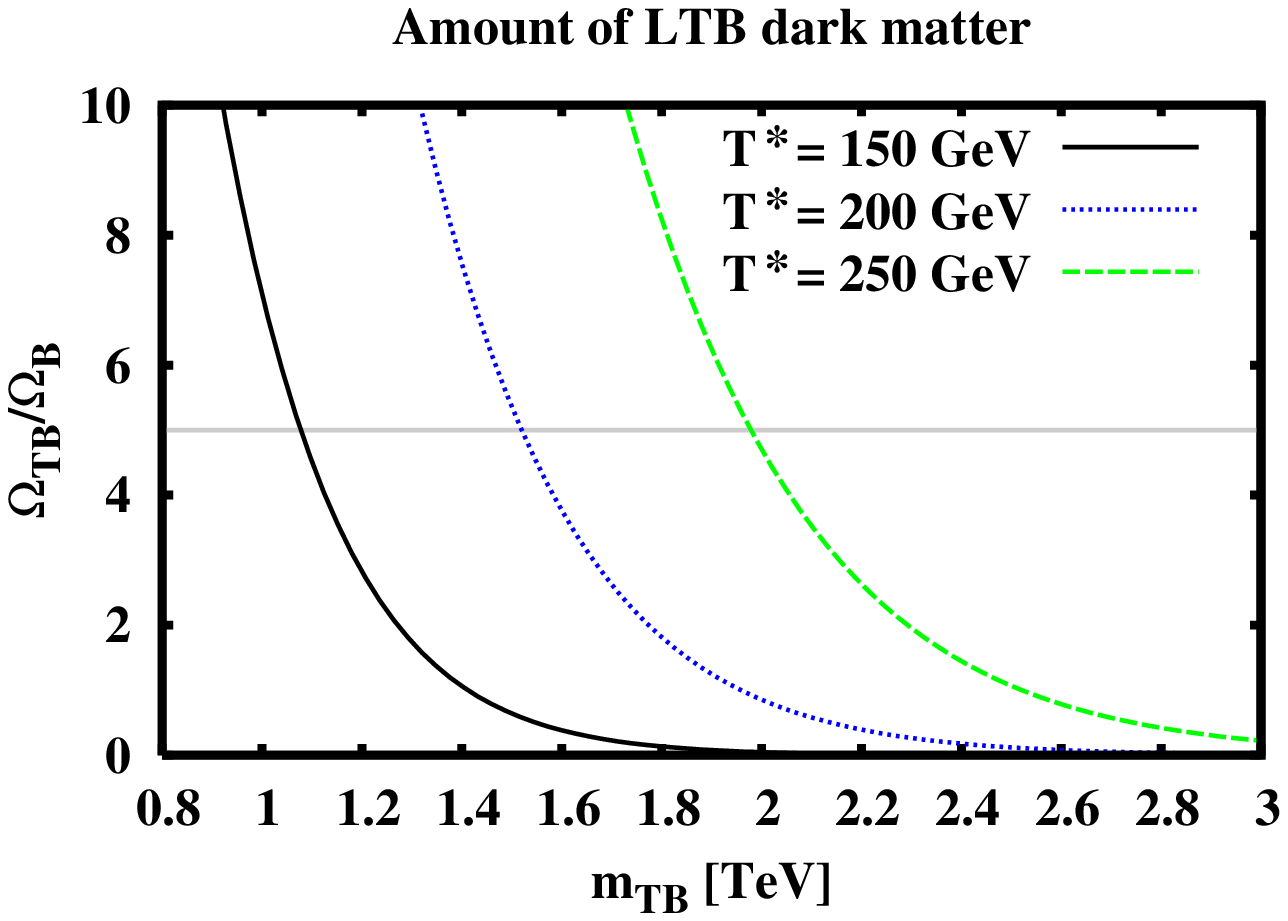}}} \quad
      \subfigure{\resizebox{!}{4.8cm}{\includegraphics{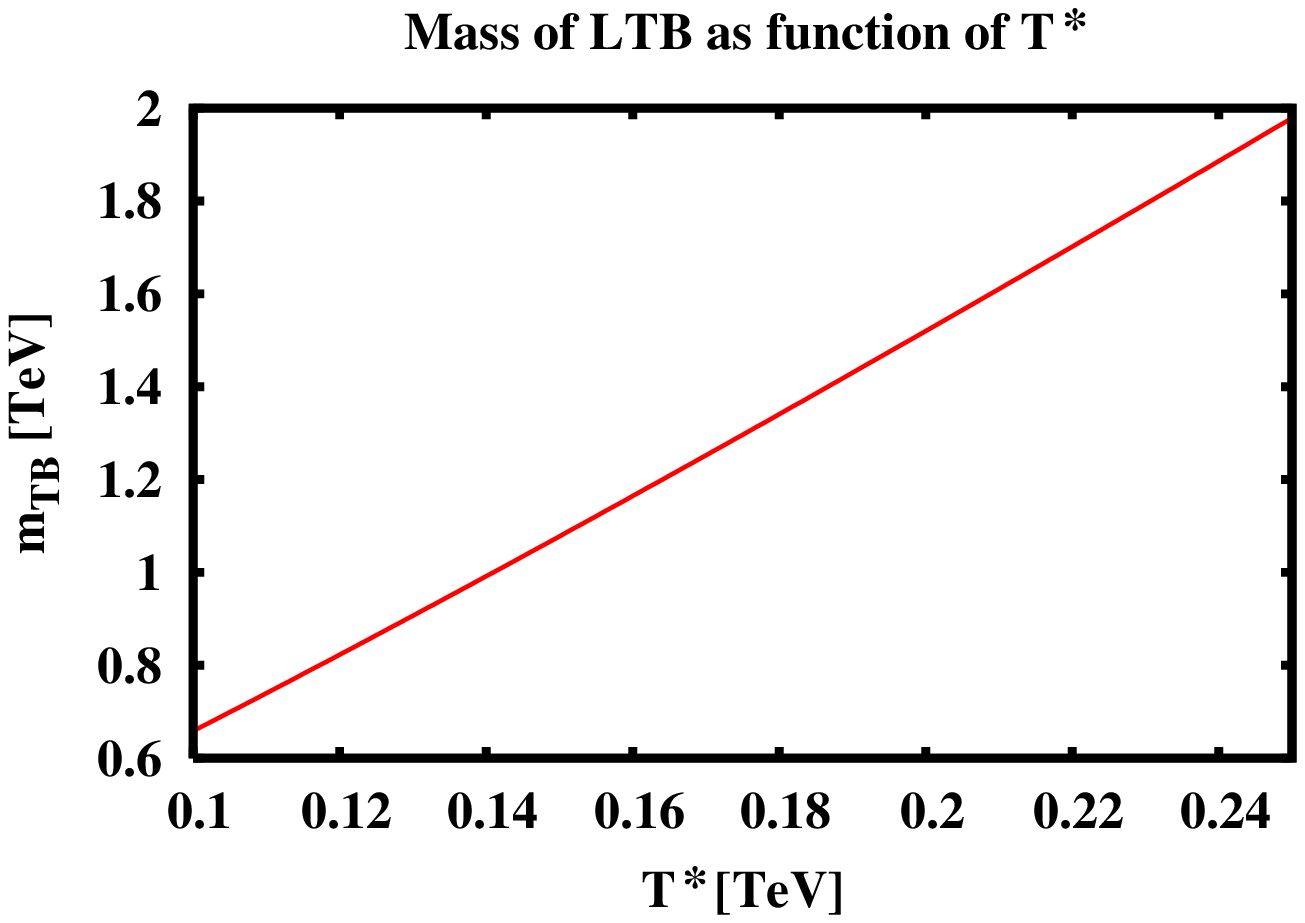}}}
      }
    \caption{\emph{Left Panel}: The fraction of technibaryon matter
    density over the baryonic one as function of the technibaryon
    mass. The desired value of $\Omega_{TB}/\Omega_{B} \sim 5$ depends
    on the lightest technibaryon mass and the value of
    $T^{\ast}$. \emph{Right Panel}: By requiring the correct amount
    ($\Omega_{TB}/\Omega_{B} \sim 5$) of dark
    matter we show the relation between the technibaryon mass and
    $T^{\ast}$.}
    \label{fig:Omega}
  \end{center}
\end{figure}

According to the choice of the hypercharge there are various
possibilities for providing a cold dark matter component.  Here we
choose the hypercharge assignment in such a way that one of the
pseudo Goldstone bosons (i.e. $DD$) does not carry electric charge.
The dynamics providing masses for the pseudo Goldstone bosons may be
arranged in a way that the neutral pseudo Goldstone boson is the lightest technibaryon
(LTB). If conserved by ETC interactions the technibaryon number
protects the lightest baryon from decaying. Since the mass of the
technibaryons are of the order of the electroweak scale they may
constitute interesting sources of dark matter. Some time ago in a
pioneering work Nussinov \cite{Nussinov:1985xr} suggested that, in
analogy with the ordinary baryon asymmetry in the Universe, a
technibaryon asymmetry is a natural possibility. A new contribution
to the mass of the Universe then emerges due to the presence of the
LTB. It is useful to compare the fraction of the total technibaryon mass
$\Omega_{TB}$ to the total baryon mass $\Omega_{B}$ in the universe
\begin{eqnarray}
\frac{\Omega_{TB}}{\Omega_B} = \frac{TB}{B} \, \frac{m_{TB}}{m_p} \
, \label{dmamount}
\end{eqnarray}
where $m_p$ is the proton mass, $m_{TB}$ is the mass of the LTB.
$TB$ and $B$ are the technibaryon and baryon number densities,
respectively.
In order to determine few features of our LTB particle, we made in
\cite{Gudnason:2006ug} the oversimplified approximation in which our
LTB constitutes the whole dark matter of
the Universe. In this limit the previous ratio should be around $5$. The fact that it is charged under $SU(2)_L$
makes it detectable in Ge detectors \cite{Bagnasco:1993st}. The
basic results \cite{Gudnason:2006ug} are shown in
Fig.~\ref{fig:Omega}.
The desired value of the dark matter fraction in the Universe can be
obtained for a LTB mass of the order of a TeV for quite a wide range
of values of $T^{\ast}$ (which is the temperature below which the
electroweak sphaleron processes cease to be relevant). The only free
parameter in our analysis is the mass of the LTB which is ultimately
provided by ETC interactions.

\section*{Acknowledgments}
I am delighted to thank D.D. Dietrich,  N. Evans, S.B. Gudnason, D.K. Hong, S.D. Hsu,  C. Kouvaris, and K.
Tuominen for having shared the work and fun on which this brief summary is based. I am indebted to T. Appelquist
for introducing me to this fascinating subject and M. Shifman for getting me interested in the problem of higher
dimensional representations. Many scientists have contributed to this subject and deserve to be cited. I hence
refer to the beautiful review of Hill and Simmons,``Strong dynamics and electroweak symmetry breaking,''
  Phys.\ Rept.\  {\bf 381}, 235 (2003)
  [Erratum-ibid.\  {\bf 390}, 553 (2004)] for a complete list of relevant references and a better
  description of some key features of dynamical electroweak symmetry breaking.
\noindent I am supported by the Marie Curie Excellence Grant as team
leader under contract MEXT-CT-2004-013510 and by the Danish Research
Agency.

\end{document}